\def\rass{\it ROSAT  All-Sky Survey\rm}
\documentclass{emulateapj}
\usepackage{amssymb,amsmath,amsthm}

\newcommand{\xmm}{\it XMM-Newton\rm}
\newcommand{\chandra}{\it Chandra\rm}
\newcommand{\rosat}{\it ROSAT\rm}

\newcommand{\euve}{\it EUVE\rm}

\newcommand{\pspc}{\it PSPC\rm}

\def\hours{$^h$}
\newcommand{\degree}{\ensuremath{^\circ}}
\def\ho{$H_0$}
\def\kmsmpc{km~s$^{-1}$~Mpc$^{-1}$}
\def\iras{\it IRAS\rm}
\shorttitle{Soft excess in Coma from RASS}
\shortauthors{Bonamente et al.}

\begin{document}


\title{The diffuse soft excess emission in the Coma cluster
from the ROSAT All-Sky Survey}


\author{M. Bonamente\altaffilmark{1,2}, R. Lieu\altaffilmark{1}, and E. Bulbul\altaffilmark{1}}

\altaffiltext{1}{Physics Department, University of Alabama in Huntsville,
Huntsville, Al 35899}
\altaffiltext{2}{NASA National Space and Technology Center, Huntsville, Al 35899}


\begin{abstract}
\rass\ data near the North Galactic Pole was analyzed  in order to study the large-scale 
distribution of soft X-ray emission from the Coma cluster.
These \rosat\ data constitute the only available X-ray observations of Coma that feature an
\it in situ \rm -- temporally and spatially contiguous -- background, 
with unlimited and continuous radial coverage. These unique characteristics of the
\rass\ data are used to deliver a
final assessment on whether the soft excess previously detected in the Coma cluster
is due to background subtraction errors, or not.
This paper confirms the presence of soft X-ray excess associated
with Coma, and reports the detection of 1/4 keV band  excess
out to 5~Mpc from the cluster center, the largest soft excess halo discovered to date.
We propose that the emission is related to filaments that converge towards Coma,
and generated either by  non-thermal radiation caused by accretion shocks,
or by thermal emission from the filaments themselves.
\end{abstract}





\section{Introduction}
X-ray emission from galaxy clusters is due primarily to a hot virialized plasma 
at a temperature $kT \sim 10^7-10^8$ keV 
that fills the intra-cluster medium (ICM). Evidence for an additional emission
component was initially discovered by \citet{lieu1996a,lieu1996b} in the
extreme ultra-violet ($h \nu \sim 0.1$ keV) with \euve\ observations
of the Virgo and Coma clusters, and then confirmed with several other instruments,
notably in the \rosat\ 1/4 keV band (see \citealt{durret2008} for a recent 
review of the literature; detections in Coma include \citealt{bowyer2004},
\citealt{finoguenov2003}, \citealt{bonamente2003}, \citealt{nevalainen2003},
\citealt{bowyer1999}).
The excess emission is usually modeled as an additional thermal component of lower temperature
($kT \sim 10^6-10^7$ keV) or as a non-thermal power law. 
An additional thermal model may arise from cooler gas inside the cluster \citep{lieu1996b},
possibly in pressure equilibrium as in the \citet{cheng2005} model, or from warm filaments
seen in projection towards clusters \citep[e.g.,][]{mittaz2004,bonamente2005}.
A non-thermal power law is indicative of relativistic electrons that scatter the
CMB photons and emit by Compton scattering \citep[e.g.,][]{sarazin1998,lieu1999}.
Given the limited spectral resolution of the current CCD 
detector technology, it has not been possible to prove conclusively which additional model
is a better fit to the data \citep[see, e.g.,][]{nevalainen2003,bonamente2005,werner2007}.

In the main X-ray band ($\sim$ 0.5-10 keV), cluster spectra in annular regions 
are typically well fit by single-temperature models.
Recently, it has become evident that the best-fit temperature
of the hot plasma is band-dependent, with a trend of lower temperatures when
lower-energy bands are used in the fit \citep{nevalainen2003,cavagnolo2008,bonamente2007};
this behavior is naturally explained by the presence of an additional soft component.

In the absence of conclusive spectral evidence on the origin of the excess emission,
clues can be found from the spatial distribution of the excess.
The excess emission typically increases with radius, with respect to the hot ICM \citep[e.g.,][]{lieu1999,bonamente2001a},
indicating that the excess component is likely unrelated to the hot ICM.
Attempts at resolving the emission have so far been unsuccessful \citep[e.g.][]{bonamente2003}, and
thus a truly diffuse origin for the excess emission is favored.
To date, the excess emission was detected out to a maximum radius of 1.7~Mpc \citep{bonamente2003},
and typically at radii $\leq$ 0.5-1 Mpc; at these radii, 
the hot plasma has a density of $\simeq 10^{-4}-10^{-5}$ cm$^{-3}$,
and can support $\sim \mu$G magnetic fields that give rise to radio halos \citep[e.g.][]{feretti2001,brunetti2007,clarke2006}.
Therefore, a non-thermal origin of the excess can be justified.

Detection of the excess emission is particularly sensitive to the background subtraction process.
The soft X-ray sky is known to have gradients on the scales of degrees \citep[e.g.,][]{snowden1997},
requiring that the background is estimated from a region that is spatially contiguous to the cluster.
Moreover, temporal variability can be induced by Solar flares, which
cause charge-exchange radiation that varies on time scales of hours; the
most accurate background subtraction is therefore performed when a simultaneous background measurement
is available, e.g., from peripheral regions of the detector.
Such time- and space-contiguous background was used in the early \euve\ and \rosat\ works,
thanks to the large field of view of those instruments. In the case of \xmm\ and \chandra, the
local background is only available for high-redshift clusters of smaller angular size.
An incorrect background subtraction caused by time variability was shown by
\citet{takei2008} to be the reason for an earlier claim of detection of O~VII emission lies associated
with the soft excess emission.
In the case of the Coma cluster, even the 1~degree radius of the \pspc\ emission is entirely filled by
cluster emission, and therefore a time-contiguous background was not available in our
earlier analysis based on pointed observation \citep{bonamente2003}.

In this paper we analyze \rass\ data in the direction of the Coma cluster.
Thanks to the all-sky nature of the observations,
these data are the only observations
of Coma to feature a background that is spatially and temporally contiguous,  and with 
unlimited radial coverage.
We therefore aim to show that excess emission 
is present when this local background is used, and determine the maximum radius of detection
of the excess emission. This paper is structured in the following way:
in Section \ref{sec:analysis} we describe the \rass\ data, the data analysis and the detection
of the soft excess emission,
in Section \ref{sec:interpretation} we interpret the emission using thermal and non-thermal models,
and in Section \ref{sec:discussion} we discuss our results and present our conclusions.

\section{Data reduction and analysis}
\label{sec:analysis}
\subsection{The \rass\ data}
\label{sec:data}
The angular resolution of the \rosat\ X-ray telescope in combination
with the \pspc\ camera is approximately 1.8\arcmin\ 
(50\% encircled energy radius averaged over the field of view),
and the energy resolution is approximately $\Delta E/E = 0.43 (E/0.93~keV)^{-0.5}$ FWHM,
corresponding to a $\Delta E \sim 0.2$ keV in the 1/4 keV  band, and
$\Delta E \sim 0.5$ keV at 1.5 keV. The field of view of \pspc\ is circular with
approximately 1 degree radius.
During the first year of operation,
\rosat\ performed an all-sky X-ray survey, which included an exposure
of $\sim 850$ seconds in the direction of the Coma cluster
\citep{briel1992}.
\citet{snowden1997} analyzed the \rosat\ All-Sky Survey (RASS) and provided
maps on the diffuse soft X-ray emission in seven bands (R1 through R7),
with linear pixel of size 12\arcmin. 
The R2 band is the softest band with reliable calibration (90\%
of peak response 0.14-0.284 keV), and band R7 (1.05-2.04 keV) is the 
highest energy band available with the \pspc\ detector. 

The maps exclude bright point-like sources identified from the RASS data themselves,
using  a flux threshold for which the RASS source catalog is complete over 90\% of the sky.
In the R1+R2 band, the flux threshold is 0.025 counts s$^{-1}$, and in the 
R6+R7 band it is 0.020 counts s$^{-1}$ \citep{snowden1997}. 
This choice resulted in the exclusion
of the central pixel of the Coma cluster in the R2 band, which appears point-like
at the resolution of these data. The units of the emission maps are $10^{-6}$ 
PSPC~counts s$^{-1}$ arcmin$^{-2}$ (a detector-specific
measure of the intensity of radiation, proportional to the
surface brightness), and take 
into account the effects of  vignetting, different exposure times
of a given sky region, detector artifacts,
obscuration by the window support structure, and other efficiency variations.
The maps of the diffuse emission are complemented by matching
maps of the estimated uncertainty in the intensity,
needed in order to determine the statistical significance of the emission
above the local background \citep{snowden1997}.~\footnote{The maps of the diffuse X-ray emission
can be found at \url{http://www.xray.mpe.mpg.de/rosat/survey/sxrb/12/ass.html}.
The R2 and R7 bands used in this work are \url{g000p90r2b120pm.fits}
and \url{g000p90s2b120pm.fits} for R2 band, and 
\url{g000p90r7b120pm.fits} and  \url{g000p90s7b120pm.fits} for R7 band.}

These X-ray data represent a unique opportunity  for the detection of diffuse emission associated
with  the Coma cluster. These are in fact the only observations that feature both a large
radial coverage and a simultaneous background measurement. The latter feature is especially
significant, in an effort to avoid
background subtraction problems
caused by the charge-exchange radiation induced
by solar activity 
\citep{takei2008}.
Owing to the scanning mode in which \rosat\ was operated during the survey, 
any enhancement caused by solar wind-induced emission will be present in both cluster
and background regions. 
No other X-ray mission has performed a survey of the X-ray sky, or has a field of view which
covers such a large area of the sky. The pointed \rass\ observations of Coma, which we
analyzed in a previous paper \citep{bonamente2003}, covered only a region of 1 degree radius,
and thus could not investigate the emission beyond that distance, or with a local background.

\subsection{Calibration accuracy of the \rosat\ PSPC}
Initial calibration of the  \rosat\ \pspc\  was  performed on the ground
\citep{pfeffermann1987}.
Of particular interest to the results of this paper is the calibration of the effective
area in the R2 band. 
\citet{snowden1995}  compared \pspc\ observations of the diffuse soft X-ray background
with observations from the Wisconsin survey \citep[e.g.][]{mccammon1983,garmire1992};
the comparison revealed that the \pspc\ effective area was well calibrated within
a 10\% uncertainty.
More recently, \citet{beuermann2006} compared the extreme-ultraviolet and soft X-ray
fluxes of three standard candles observed by the \chandra\ LETG, \euve\ and 
\rosat\ \pspc, and found that the agreement between the various instruments
indicate that the \pspc\ effective area is calibrated within a few \% error.
These results indicate that the uncertainties in the R2 band effective area
are of the order of few percent; as it will be shown in the following, this level
of uncertainty will not affect any of the results presented in this paper.

\subsection{Images and radial profiles of the X-ray surface brightness}
\label{sec:profiles}
\begin{figure}[!h]
\begin{center}
\includegraphics[width=2.5in,angle=-90]{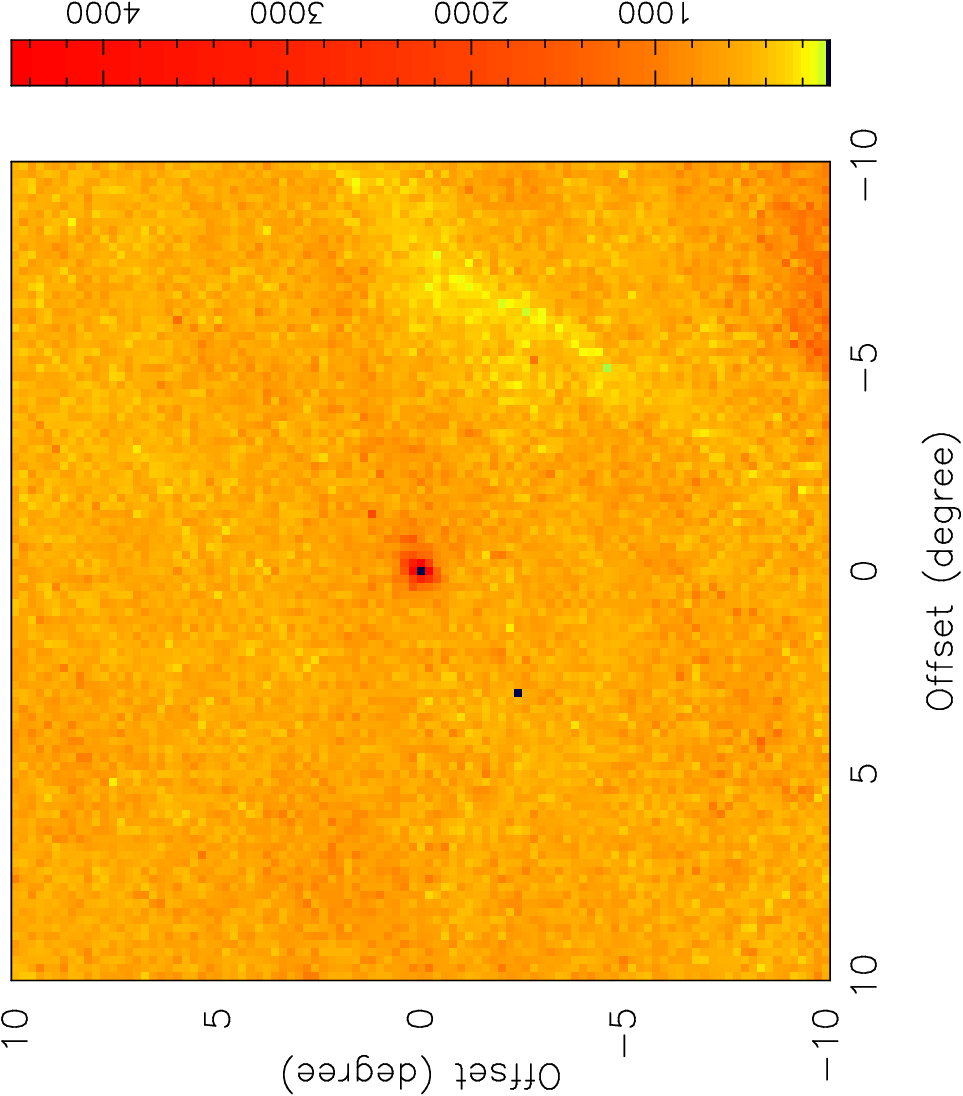}
\includegraphics[width=2.5in,angle=-90]{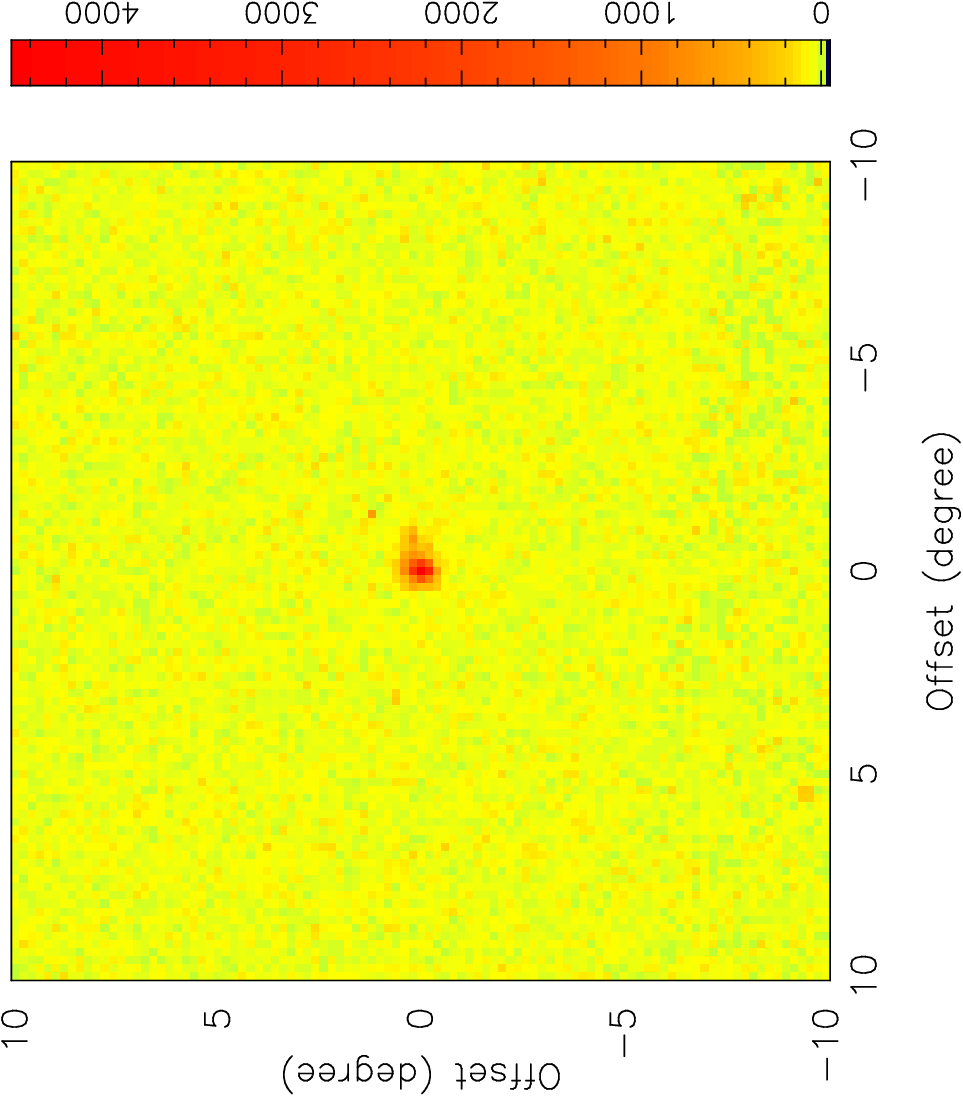}
\end{center}
\caption{Images of the diffuse emission of Coma in bands R2 (a) and R7 (b)
\label{fig:coma}. Units of the emissions are
$10^{-6}$ PSPC~counts s$^{-1}$ arcmin$^{-2}$.}
\end{figure}

In Figure \ref{fig:coma} we show the diffuse surface brightness  in a region
of 400 square degrees around the Coma cluster. The R2 band map features the
northern edge of the North Polar Spur in the bottom right quadrant of panel (a);
the two black pixels indicate the location of two
point-like features removed by the \citet{snowden1997} data reduction.
The R7 band image indicates that the 1-2 keV local background in the neighborhood of Coma
does not feature any large-scale gradients.

\begin{figure}[!h]
\begin{center}
\includegraphics[angle=-90,width=3.0in]{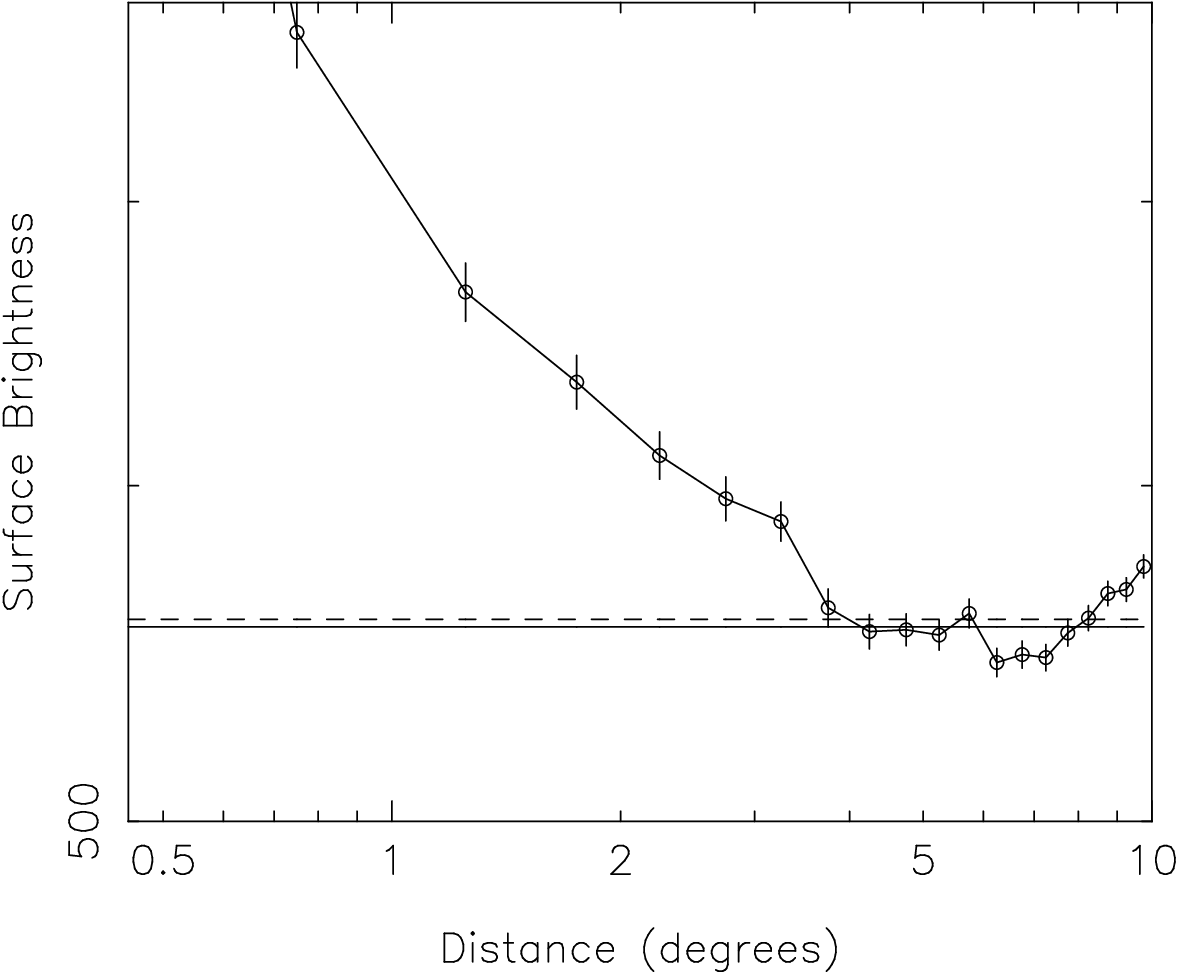}
\includegraphics[angle=-90,width=3.0in]{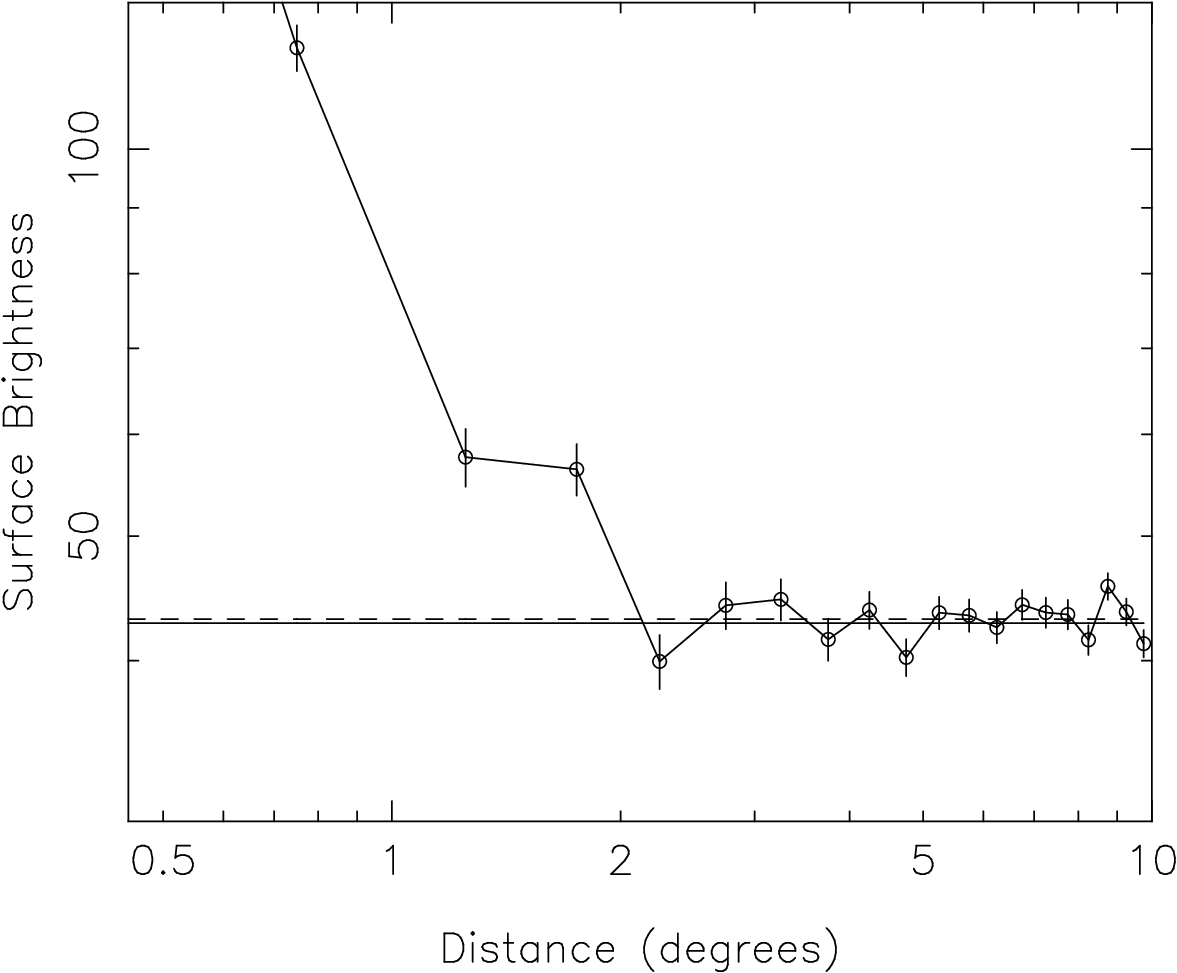}
\end{center}
\caption{Radial profiles of the diffuse emission near Coma in R2
(a) and R7 band (b). Units of the surface brightness are  \label{fig:radial-profiles}
$10^{-6}$
PSPC~counts s$^{-1}$ arcmin$^{-2}$.}
\end{figure}

We obtained azimutally averaged radial profiles of the surface brightness
in the two bands (Figure \ref{fig:radial-profiles}), centered at R.A=12\hours $59\arcmin 50\arcsec$,
Dec.=$+27\degree 58\arcmin 59 \arcsec$, located at the center of
Figure \ref{fig:coma}.~\footnote{This position 
corresponding to pixel (247.80,251.08)
in the Snowden maps.} 
Analysis of the R2 profile  in Figure \ref{fig:radial-profiles}(a) shows that
the North Polar Spur region is likely to affect the level of emission at distances
$\geq 6\degree$ from Coma. We therefore estimate the local background in two ways:
as the average of the $4-6\degree$ region, and as the average of the $4-10\degree$ region
(Table \ref{tab:background}). Given that the two estimates of the background are statistically
consistent with one another in each band, we use the $4-6\degree$ background in the following
analysis; all results presented
in this paper are unchanged if the $4-10\degree$ background is used instead.
The choice of $4\degree$ as the minimum distance is motivated by the fact that the
emission appears to fade into the background between 3 and 4 degrees, consistent
with the fact that Coma is not known to emit X-rays at such large distances.
For a Hubble constant of \ho=73~\kmsmpc, 1 degree corresponds to 1.67 Mpc at the redshift
of the Coma cluster ($z=0.024$).

The radial profiles indicate  the 1-2 keV emission reaches the local background at a radius
of $\sim 2\degree$, while the softer R2 band has emission above the local background out to
(at least) $3\degree$. The radial profiles therefore indicate that the Coma cluster has
an additional soft emission component at large radii.
Measurements of the virial radius of Coma were recently obtained by
weak lensing measurements (\citealt{kubo2007}, $r_{200}=2.7\pm0.3$ Mpc
for $H_0=73$ \kmsmpc), galaxy kinematics (\citealt{lokas2003},
$r_{v}=2.8\pm0.8$ Mpc for $H_0=73$ \kmsmpc) and redshift survey of cluster
galaxies  (\citealt{geller1999}, $r_{200} \simeq 2.1$ Mpc for $H_0=73$ \kmsmpc).
These measurements are in agreement with earlier X-ray measurements
by \citet[based on these \rosat\ data]{briel1992} and \citet[based
on \it Einstein \rm data]{hughes1989},
%(the latter based
%on \it Einstein \rm data)
which detected 
X-ray emission at radial distance of $\leq$100 arcmin.
The present detection of soft X-ray emission out to (at least) 3\degree\,
or 5 Mpc, therefore tracks the soft X-ray emission out to 
the virial radius, and   also constitutes the largest continuous 
halo of X-ray emission
detected to date in any galaxy cluster.

\subsection{The soft X-ray excess emission}
\label{sec:excess}
Futher analysis of the soft emission must take into account the contribution
of the hot intra-cluster medium to the R2 band emission. Given the  limited spectral resolution
and narrow bandpass ($\sim$0.2-2 keV), the \rass\ \pspc\ data are  not ideal for the determination of hot plasma
temperatures. We therefore rely on other X-ray studies of Coma, which consistently measure
a plasma temperature of $kT \simeq 8$ keV in the central $20\arcmin$ 
\citep[e.g.,][]{arnaud2001,lutovinov2008}.
At larger radii, \citet{finoguenov2001} measure temperatures in the
range of 15--3 keV  for regions
between 0.5-1.5$\degree$ from the cluster center, 
consistent with the decrease of temperature at large
radii observed in other clusters \citep{vikhlinin2006}.

In order to estimate the hot ICM contribution to the R2 band, we assume that the average plasma
temperature has the distribution shown in Table \ref{tab:t}, decreasing from 8~keV to 2~keV
between the center and the outskirts.  We estimate the ratio of
R2-to-R7 band count rates using the method described in \citet{snowden1997}, which consists
of using the on-axis \pspc\ response function in conjunction with an optically thin
plasma emission model~\footnote{This procedure was also confirmed by S. Snowden, private communication.
The \pspc-C camera was used for the All-Sky Survey.}.
The distribution of Galactic HI in the direction of Coma was investigated in \citet{bonamente2003},
by means of the \citet{dickey1990} and \citet{hartmann1997} 21-cm data, and \iras\ 100 $\mu$m data.
For a region within 5\degree\ from the cluster center, the HI column density is between
$N_H=0.8-1.1 \times 10^{20}$ cm$^{-2}$, with no evidence of large-scale gradients;
for the count-rate ratios in Table \ref{tab:t}, we assumed $N_H=0.9 \times 10^{20}$ cm$^{-2}$, which
is appropriate for the radial range of interest.
The count-rate ratio is also sensitive to the chemical composition of the plasma; in particular,
a higher abundance $A$ of metals results in a lower R2-to-R7 ratio, due to emission lines
processes. We therefore used a conservative value of $A=0$ in deriving the  estimates of
Table \ref{tab:t}; if an abundance of $A=0.3$ was used instead, the count-rate ratio
would decrease respectively to 1.03 (8 keV), 1.20 (4 keV) and 1.48 (2 keV), causing
a lower prediction for the contribution of the hot ICM in the R2 band and higher soft
X-ray fluxes.
For each annulus, we thus estimate the hot ICM contribution to the 1/4 keV band 
by multiplication of  the background-subtracted R7 intensity (Figure \ref{fig:radial-profiles}) by the
factor in Table \ref{tab:t}.

\begin{table}[!t]
\begin{center}
\begin{tabular}{l|cc}
\hline
 & R2 & R7 \\
\hline
$4-6\degree$ & $555.7 \pm 2.3$ & $42.8 \pm 0.7$ \\
$4-10\degree$ & $558.0 \pm 1.2$ &$43.1 \pm 0.3$ \\
\hline
\end{tabular}
\end{center}
\caption{Background levels in $10^{-6}$ PSPC~counts s$^{-1}$ arcmin$^{-2}$. 
\label{tab:background}}
\end{table}

\begin{table}[!t]
\begin{center}
\begin{tabular}{l|cc}
\hline
Region & $kT$ &R2/R7 \\
(\degree) & (keV) &  c/r ratio\\
\hline
0-0.5 & 8 & 1.04 \\
0.5-2 & 4 & 1.28\\
2-4 & 2 & 1.80 \\
\hline
\end{tabular}
\end{center}
\caption{Average temperature of the Coma plasma, and ratio of R2-to-R7 count rate for 
\pspc.\label{tab:t}}
\end{table}

\begin{figure}[!h]
\begin{center}
\includegraphics[angle=-90,width=3in]{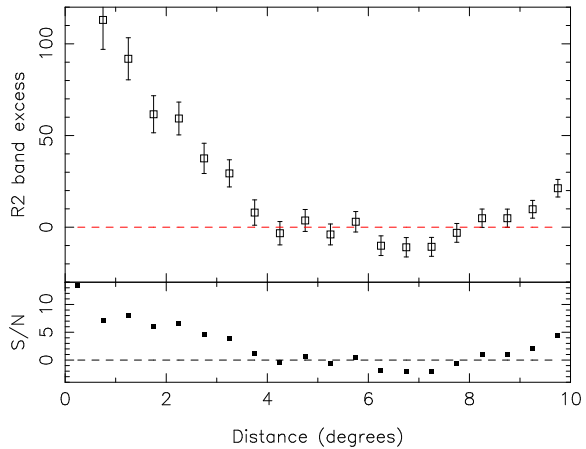}
\end{center}
\caption{Radial profile of the excess emission in R2 band, above the contribution
from the hot ICM \label{fig:excess}. Units of the excess emissions are
$10^{-6}$ PSPC~counts s$^{-1}$ arcmin$^{-2}$.}
\end{figure}

In Figure \ref{fig:excess} we show the R2 band excess emission above the contribution of the hot ICM.
This determination of the excess emission clearly depends on our choices in the
 modelling of the thermal plasma. If the plasma temperature
in the 0.5-2\degree\ region is higher, then our estimates are a strict lower limit. 
The excess emission is still detected at high significance
by using a temperature as low as 2 keV for the 0.5-2\degree\ region;
%%%%%%%%%
in this case, the significance of detection of the three bins in Figure \ref{fig:excess} 
between 0.5 and 2\degree\ decreases,
respectively, from 7.0, 8.0 and 6.1 $\sigma$ to 4.2, 6.9 and 5.1 $\sigma$.
%%%%%%%%%
The uncertainty in the R2 excess takes into account the error in the background measurements from the
4-6\degree\ region. As noted above in Section \ref{sec:profiles}, the excess is detected
with same statistical significance if the 4-10\degree\ region is used to estimate the background.
%%%%%%%%% 
The increase in the  1/4 keV band flux at radii $\geq 8$ \degree\ is associated with
a North Polar Spur feature clearly visible in Figure \ref{fig:coma}, and gives rise
to a spurious 1/4 keV excess. This emission was the reason for our choice of the
local background in the 4-6\degree\ region -- which is less affected by the North Polar Spur
emission-- and highlights the need of a local background for the purpose of background
subtraction.
%%%%%%%%%

\section{Interpretation of the excess emission}
\label{sec:interpretation}
In Section \ref{sec:analysis} we discussed possible sources of systematic
error that may affect the excess emission shown in Figure \ref{fig:excess}, and concluded
that the signal cannot be explained by errors in the HI column density, modelling of the hot plasma,
or background subtraction. We therefore proceed to investigate if the emission can be due to unresolved point
sources, and then discuss two scenarios for a cluster origin of the emission.

\subsection{Point source emission}
Given the limited angular resolution of these \rass\ data, a possible 
explanation for the detected excess is that the signal is associated with a number
of unresolved point sources (e.g., galaxies) in the Coma field. 
We address this possibility by using the analysis of
X-ray point sources detected by \citet{finoguenov2004} in an \xmm\  mosaic observation of the 
central Coma cluster with the \xmm\ EPIC detector. 
Their analysis detects a number of X-ray point sources for a total
combined flux of $F=0.67\pm0.01$ EPIC~counts per second in the 1-2 keV
band, in a circular area of approximate 
radius $0.78\degree$. We can use these numbers to estimate the contribution of unresolved point
sources in our \rass\ data. First, we rescale the EPIC count rate to the PSPC count rate in the
same energy band, taking into account that the PSPC camera has an average effective
area that is $\sim$~7.5\% that of the EPIC detectors, in the 1-2 keV band \citep{finoguenov2004,
snowden1997}. Then, assuming (conservatively) that an equivalent population of X-ray sources is present at larger
radii, we can estimate that the average brightness due to point sources in the R7 band is of order
$S=7.2\times 10^{-6}$ PSPC~counts per second per square arcmin.
The contribution of these point sources 
to the R2 band depends on their spectrum; for an  X-ray
thermal spectrum with $kT=1$ keV, typical of galaxies,
 the R2 count rate is approximately 2-2.5 times that in the R7 band,
i.e., $\sim 1 \times$  higher than the R2-to-R7 conversion factor used at radii $\geq 0.5 \degree$
in converting the R7 emission to the R2 band (see Table \ref{tab:t}). 
Unresolved point sources  would therefore reduce the soft excess fluxes
of Figure \ref{fig:excess} by approximately $7.2\times 10^{-6}$ PSPC~counts per second per square arcmin,
which corresponds to 10\% of the detected excess at radii less than 2\degree, and by 10-20\% in the
2-3\degree\ region. We therefore conclude that unresolved point source
do not contribute significantly, if at all,  to the detected signal.

\subsection{Diffuse excess emission associated with the Coma cluster}
The only remaining possibility is that the detected signal is a truly diffuse celestial component
related to the Coma cluster. These data therefore constitute the detection of soft X-ray excess
emission out to a radial distance of approximately 3\degree, or 5 Mpc, from the center of
the Coma cluster. The main feature of the excess emission is its persistence \it beyond \rm the
radius where the hot ICM is detected ($\leq 2\degree$), thus providing strong evidence of the
fact that the excess component is unrelated to the thermal plasma.
We propose two possible explanations for this large-scale emission.

The first explanation is that the emission is due to Compton scattering of
CMB photons off of relativistic electrons, which have been accelerated at 
locations where gas accreting into the cluster potential becomes supersonic,
thus exciting shock waves. The radio observations of \citet{bagchi2006} show
a very spectacular detection of a ring of emission at a distance of 
$\sim 1$ Mpc from the center of  the cluster Abell~3376,
which is interpreted as the result of accretion shocks. 
There is theoretical \citep[e.g.,][]{cen1999,dave2001,dolag2006} and observational evidence
\citep[e.g.,][]{finoguenov2003,werner2008} that diffuse filaments at $kT\simeq 10^5-10^7$ K
are present in the inter-cluster medium; these structures are therefore the 
ideal candidate for providing gas accreting into the deeper cluster potential.
Shock acceleration, requires the presence of magnetic fields, in order to generate
Alfv\`{e}n waves that act as scattering centers \citep[e.g.][]{bell1978a}.
Magnetic fields are clearly present in  Abell~3376, as revealed by the non-thermal
radio emission; it is however not clear, though, how magnetic fields would arise
in an environment where the density of gas is expected to be very low ($\leq 10^{-5}$ cm$^{-3}$).

Another interpretation is that optically-thin soft X-ray emission from the filaments themselves
is responsible for the detected emission. This scenario has been discussed in several prior
papers \citep[e.g.,][]{bonamente2003,mittaz2004,bonamente2005}, in which it was found that the
column density of the filaments would exceed the expectations based on numerical simulations.
We use the present \rass\ data to estimate the characteristics of the filaments that may be
responsible for the excess emission. We assume filaments at a temperature of $kT=0.1$ keV,
and with null abundance of metals, typical of intergalactic filaments, and use the
APEC emission code \citep{smith2001} to  calculate the emissivity as $\Lambda=4.5\times 10^{-16}$
counts cm$^3$ s$^{-1}$ in the 0.14-0.284 keV (R2) band.
We also assume a uniform electron density of $n=10^{-4}$ cm$^{-3}$,
corresponding to a baryonic overdensity of $\delta \sim 200-300$ 
(for \ho=73 \kmsmpc),  an average filament
length of $L=5$ Mpc in the direction of the observer. Using 
the average R2 band effective area of the \pspc\ detector
($A_{eff} \simeq 150$ cm$^2$) the R2 band surface brightness is calculated as
\begin{equation}
S_X = \frac{1}{4 \pi} \Lambda \cdot n^2 L \cdot  A_{eff} =70 \times 10^{-6} \hspace{0.3cm} 
\text{(counts s$^{-1}$ arcmin$^{-2}$)}
\end{equation}
Comparison with Figure \ref{fig:excess} shows that this is in fact the typical excess emission detected
at the outskirts of the Coma cluster (radii $\gtrsim 1.5 \degree$). 
Since the surface brightness is proportional to $n^2 L$,
lower density filaments would require to be substantially longer along the sightline,
in order to explain the detected emission.
Filaments of this density or length are more massive than typical filaments in
numerical simulations \citep[e.g.,][]{mittaz2004}. A possible explanation for the variance
between observations and numerical models 
is that the filaments are actually magnified by gravitational lensing caused by the
cluster potential, as discussed in Lieu and Bonamente (2009, submitted).

An additional interpretation can be provided by the \citet{prokhorov2008} model 
in which the soft excess emission is the result of a non-equilibrium state between
electrons and ions present at the cluster's outskirts.
The model predicts soft X-ray excess from low-temperature electrons
near the virial radius, which is approximately 2-3\degree\ (see Section \ref{sec:profiles}).
These \rass\ observations are compatible with such scenario.
\section{Discussion and conclusions}
\label{sec:discussion}
These \rass\ data provide a unique view of the soft excess emission from the Coma cluster, which
is now detected out to a $\sim$5 Mpc radius.
Diffuse filaments of warm gas provide the natural interpretation for the excess emission,
either via particle acceleration at accretion shocks, or simply via their thermal emission.
The scenario of thermal emission from filaments was tested using \xmm\ observations of
Abell~S1101 \citep{bonamente2005}; in that case, we determined that the filaments would 
be required to be even more massive than the ones discussed in this paper (for the same nominal
density $n=10^{-4}$ cm$^{-3}$, the filament length would exceed 10~Mpc).
In an upcoming paper (in preparation), we will show that the excess emission 
in Abell~S1101 was overestimated, due to a significant 
revision of the HI column density in the direction to the cluster \citep{kalberla2005}; 
the reduced flux will make the estimates similar to the ones presented in this paper.

Current X-ray missions are not designed for studies of large-scale soft X-ray emission, given
their narrow field of view and issues in the calibration of the 1/4 keV band (see
\citealt{nevalainen2007} for a discussion on the \xmm\ soft band, and \citealt{bonamente2007}
on \chandra).
The presence of filaments has therefore been mainly investigated in absorption, giving
few tentative detections of absorption lines due to filaments.
In particular, \citet{nicastro2005} detected $z>0$ absorption lines towards Markarian~421 with \chandra, although
\citet{rasmussen2007} did not confirm the detection with \xmm.
Of particular relevance to the Coma excess is the work of \citet{takei2007}, 
who used \xmm\ RGS data to obtain a 3-$\sigma$ detection 
of absorption lines in the spectrum of X Comae, a background quasar at a projected
distance $\sim 25\arcmin$ from the cluster center. Although that detection is of limited statistical
significance, it provides support to the thermal interpretation of the soft excess
emission we discussed in Section \ref{sec:excess}, at least in the central regions of the cluster. 

The \rosat\ mission continues to provide the most compelling detections of soft excess emission from galaxy
clusters. Using pointed \pspc\ data, we were initially able to detect soft excess emission in several
clusters \citep{bonamente2001a,bonamente2001b,bonamente2001c}, confirming the
discovery papers based on \euve\ observation \citep{lieu1996a,lieu1996b}.
We then
showed that a large fraction of clusters at high
Galactic latitude feature the excess emission \citep{bonamente2002},
used a mosaic of four pointed observations of Coma to detect the emission out to 2.6~Mpc
\citep{bonamente2003} and now, based on the \rass\ data, out to 5~Mpc.
In all cases, \rosat's unique combination of a reliable calibration in the 1/4 keV band, and the
availability of contemporaneous background, rendered these detections possible.
The main advantage of searching for filaments in emission is the insensitivity to the
abundance of metals, and to clumping of the gas. The observations analyzed in this paper show that
a very modest exposure time with a wide-field soft X-ray camera is a most efficient way to
detect emission from warm filaments in the neighborhood of clusters.

\acknowledgments
We are grateful to Dr. S. Snowden for comments and suggestions on the data analysis.
%\email{aastex-help@aas.org}.


%% in parentheses, after the keyword, but they will not be verified.

%{\it Facilities:} \facility{Nickel}, \facility{HST (STIS)}, \facility{CXO (ASIS)}.



%\appendix

%\section{Appendix material}

%Consider once again a task that computes profile parameters for a modified

\newpage
%\bibliography{coma}
%\bibliographystyle{apj}

%\begin{thebibliography}{}
%\bibitem[Auri\`ere(1982)]{aur82} Auri\`ere, M.  1982, \aap,
%    109, 301
%\end{thebibliography}

\clearpage

%% the file, uncomment the \plotone, \plottwo, and
%% If you need a layout that cannot be achieved with \plotone or
%% \plottwo, you can invoke the graphicx package directly with the
%% \includegraphics command or use \plotfiddle. For more information,
%% please see the tutorial on "Using Electronic Art with AASTeX" in the
%% documentation section at the AASTeX Web site,
%% http://www.journals.uchicago.edu/AAS/AASTeX.
%%
%% The examples below also include sample markup for submission of
%% supplemental electronic materials. As always, be sure to check
%% the instructions to authors for the journal you are submitting to
%% for specific submissions guidelines as they vary from

%% This example uses \plotone to include an EPS file scaled to
%% 80% of its natural size with \epsscale. Its caption
%% has been written to indicate that additional figure parts will be
%% available in the electronic journal.

\end{document}